\documentclass[10pt, conference]{IEEEtran}
\IEEEoverridecommandlockouts
\pdfoutput=1
\usepackage{cite}
\usepackage{amsmath,amssymb,amsfonts}
\usepackage{graphicx}
\usepackage{textcomp}
\usepackage{xcolor}
\usepackage{soul}

\usepackage{graphicx}
\usepackage{stfloats}
\usepackage{url}
\usepackage{color}
\usepackage{multicol}
\usepackage{booktabs}
\usepackage{balance}
\usepackage{cancel}
\usepackage{comment}
\usepackage{amsmath}
\usepackage{amsfonts}
\usepackage{subfigure}
\usepackage{amssymb}
\usepackage[shortcuts]{extdash} 

\def\BibTeX{{\rm B\kern-.05em{\sc i\kern-.025em b}\kern-.08em
    T\kern-.1667em\lower.7ex\hbox{E}\kern-.125emX}}
\begin{document}

\title{AdEle: An \underline{Ad}aptive Congestion-and-Energy-Aware \underline{Ele}vator Selection for Partially Connected 3D NoCs}

\author{\IEEEauthorblockN{Ebadollah Taheri, Ryan G. Kim, and Mahdi Nikdast}\\ \vspace{-0.4cm}
\IEEEauthorblockA{Department of Electrical and Computer Engineering, Colorado State University, Fort Collins, CO 80523, USA}
\vspace{-2.1em}
}

\vspace{-1in}

%\IEEEoverridecommandlockouts
%\IEEEpubid{\makebox[\columnwidth]{978-1-6654-3274-0/21/\$31.00~\copyright2021 IEEE \hfill} \hspace{\columnsep}\makebox[\columnwidth]{ }}

\maketitle

\begin{abstract}
By lowering the number of vertical connections in fully connected 3D networks-on-chip~(NoCs), partially connected 3D NoCs~(PC\=/3DNoCs) help alleviate reliability and fabrication issues. This paper proposes a novel, \underline{ad}aptive congestion- and energy-aware \underline{ele}vator-selection scheme called AdEle to improve the traffic distribution in PC\=/3DNoCs. AdEle employs an offline multi-objective simulated-annealing-based algorithm to find good elevator subsets and an online elevator selection policy to enhance elevator selection during routing. Compared to the state-of-the-art techniques under different real-application traffics and configuration scenarios, AdEle improves the network latency by 10.9\% on average (up to 14.6\%) with less than 6.9\% energy consumption overhead.
\end{abstract}
%improvement in real application (against CDA): PS1=9.3  PS2=14%  PS3=21.4

\begin{IEEEkeywords}
Partially connected 3D networks-on-chip, through-silicon via, simulated annealing, elevator selection.
\end{IEEEkeywords}

\pdfoutput=1
\section{Introduction}
%3D NoCs benefits
% 1- Network: High connectivity and low average internode distance 
% 2- Compatible with 3D IC --> Power efficient + Heterogeneous integration: Can realize different requirements in different layers
Network-on-chip~(NoC) has become the prevailing solution to enable scalable on-chip communication in manycore systems. Moreover, with the advances in three-dimensional~(3D) integration technologies, 3D NoCs are emerging to further improve the heterogeneity and integration density by vertically stacking multiple dies connected with an efficient die-to-die interconnect~\cite{arka2020making}. Among different vertical interconnect technologies, through-silicon vias~(TSVs) promise high bandwidth and low power~\cite{lu2017tsv, salamat2018lead, coelho2019fl}.

% Why Partially Connected topology?
% 1- TVS faults makes topology partially connected
%   1-a-Main sources of faults
%   1-b-High fault rate of TSVs
% 2- To improve fabrication cost
TSV-based 3D NoCs have been proposed for various applications~(e.g., \cite{wang2017hrc, vu2019fault}). However, vertical links in TSV-based 3D NoCs use multiple TSVs in a bundle, resulting in high area overhead due to the large TSV interconnect-pitch and keep-out-zone requirements~\cite{wang2014effective}. Also, TSVs are particularly susceptible to electromigration and capacitive crosstalk-induced issues~\cite{frank2011resistance, eghbal2015analytical}. Therefore, 3D NoC architectures with TSVs at every router (i.e., fully connected) impose higher design complexity, fabrication costs, and performance degradation~\cite{coelho2019fl, arka2020making}. Addressing such challenges has motivated the development of 3D NoCs with fewer TSV-based vertical links, also known as partially connected 3D NoCs~(PC\=/3DNoCs)~\cite{arka2020making, dubois2011elevator,taheri2020addressing}. 

% Latency and energy efficiency challenges
% 1- High traffic pressure on elevators --> impose latency overheads
% 2- Non-minimal routing in irregular topologies --> impose energy overhead 
Nevertheless, PC\=/3DNoCs introduce some new design challenges because of their partial vertical connectivity~\cite{fu2019congestion, taheri2020addressing}. In particular, the vertical links (a.k.a. the elevators) must be shared among multiple routers, potentially creating traffic hotspots at the elevators and increasing the network latency~\cite{arka2020making}. To balance the traffic at these hotspot elevators, an adaptive routing technique is needed to select lower utilized elevators without detouring far from the minimal path (i.e., elevator-selection problem). Yet, initial routing solutions in PC\=/3DNoCs (e.g., Elevator\=/First routing ~\cite{dubois2011elevator}) na\"ively select the nearest elevator without considering the traffic, resulting in unbalanced elevator utilization. To reduce congestion, advanced methods~(e.g., CDA~\cite{fu2019congestion}) use global traffic information to improve the traffic distribution during runtime. However, retrieving global traffic information increases both the hardware overhead and network traffic.

% Paper contribution
% 1- Discuss elevator assignment concerns
% 2- Propose Amun
%    2-a- Employ AMOSA for optimization
%    2-b- Add a dynamic policy to improve assignment
This paper addresses the elevator-selection problem in PC\=/3DNoC routing techniques by developing, for the first time, a novel, congestion- and energy-aware \underline{ad}aptive \underline{ele}vator-selection scheme called AdEle. AdEle works in two stages to balance the traffic with minimal overhead: an offline elevator-set optimization and an online elevator-selection policy. In the offline elevator-set optimization, AdEle uses a multi-objective simulated-annealing-based optimization algorithm (AMOSA~\cite{bandyopadhyay2008simulated}) to collectively choose an optimized subset of elevators for each source router that minimizes the average latency and energy under an assumed traffic scenario. During runtime, each router monitors its local traffic and selects one elevator from its subset to improve the latency of the network. AdEle employs a low-overhead local traffic monitoring technique that examines the blocking as a proxy for path congestion, balancing the elevator traffic while eliminating the overhead of global traffic monitoring used in other approaches. Our results simulated using different real-application traffics and configuration scenarios show the promise of AdEle compared to the state-of-the-art techniques: on average, AdEle improves the network latency by 10.9\% (up to 14.6\%) and with only 6\% (up to 6.9\%) energy consumption overhead.

% Paper organization
The rest of the paper is organized as follows. We review the recent related work on PC\=/3DNoCs in Section~II. Section~III discusses the elevator-selection problem and its complexity in PC\=/3DNoCs and details our proposed technique and its implementation. We present our simulation results in Section~IV. Finally, Section~V concludes the paper.
\pdfoutput=1
\section{Background and Related Work}
%Routing in Partially Connected NoCs 
Employing conventional dimension-order routing algorithms in PC\=/3DNoCs will result in deadlock because of the irregular topology in such networks. To prevent deadlock, the Elevator\=/First routing algorithm~\cite{dubois2011elevator} employs two virtual networks to break cyclic dependencies. Moreover, as the elevator-less routers cannot directly send packets to other layers, an elevator is selected for each packet to facilitate the inter-layer communication. Leveraging such a principle, several routing algorithms have been proposed for PC\=/3DNoCs~\cite{salamat2018lead, lee2015redelf}. However, they follow an elevator-selection policy that ignores elevators' load distribution and the minimal path. This can be especially harmful for PC\=/3DNoCs with non-uniform elevator placements, small number of elevators, or non-uniform traffic distributions. Adaptive elevator-selection techniques have been proposed~\cite{salamat2016resilient,coelho2019fl,taheri2020addressing} but mainly focus on elevator failure concerns. These strategies select the closest non-faulty elevator to the source without considering the elevator's congestion, causing them to suffer from high energy and latency costs. 
\begin{figure}[t]
\centering
\includegraphics[scale=0.78]{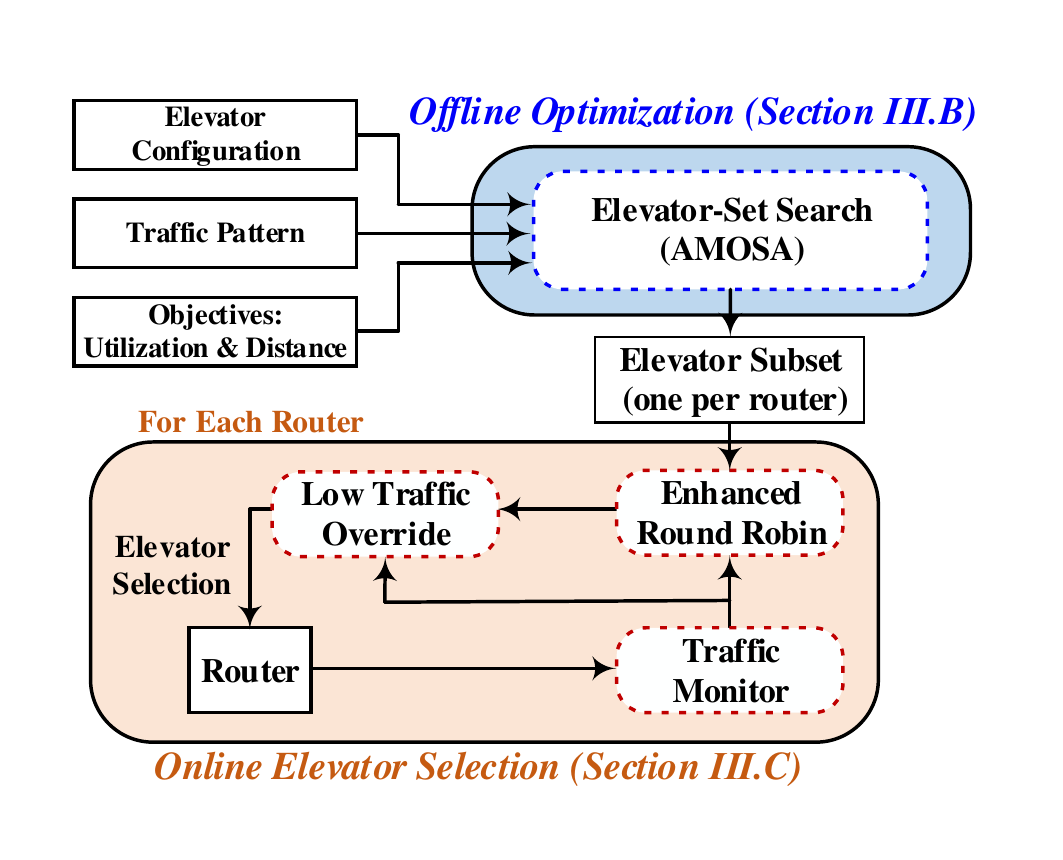}
\vspace{-0.1in}
\caption{An overview of our proposed elevator-selection scheme: AdEle.}
\vspace{-0.20in}
\label{overview_fig}
\end{figure}

To improve the traffic distribution in PC\=/3DNoCs, \cite{foroutan2014assignment} proposed an optimized elevator-selection scheme using the Tabu search algorithm. However, the offline Tabu optimization cannot capture the dynamics of the runtime network traffic.  Also, the search algorithm ignores the network energy efficiency during the elevator selection. In~\cite{fu2019congestion}, an online elevator-selection scheme called CDA selects the elevator based on the buffer utilization of the routers between a source and the elevator. However, CDA requires online global information of the network buffer utilization which imposes high latency and hardware overheads to share this information.

Considering the aforementioned works, an efficient elevator-selection solution is essential but yet to be addressed for PC\=/3DNoCs. We take on this challenge by developing a novel, adaptive congestion- and energy-aware elevator-selection scheme (AdEle). Offline elevator-selection approaches enjoy low overhead while online approaches achieve better network latency and energy consumption. Accordingly, AdEle combines the benefits of both approaches while also considering energy consumption. On top of being energy-aware, AdEle includes elevator redundancy and online policies to accommodate dynamic traffic behavior. We will show that using a set of elevators instead of one elevator for each router can greatly improve network performance. Also, our proposed approach only utilizes local information of routers to effectively manage elevator congestion with low overheads.
\vspace{-0.01in}

%However, Elevator\=/First routing is deterministic and cannot efficiently manage runtime traffic congestion. The adaptive routing algorithm LEAD~\cite{salamat2018lead} improves upon the Elevator\=/First by providing higher intra-layer path diversity. Yet, the elevator (inter-layer path) selection is not adaptive in LEAD, and hence it fails to consider and overcome traffic congestion in the elevators. 
%\input{3-RoutingInVPC}
%\input{OldVersionSEction3}
\pdfoutput=1
\section{Proposed Elevator-Selection Scheme: AdEle}
This section details our proposed adaptive congestion- and energy-aware elevator-selection scheme. As shown in Fig.~\ref{overview_fig}, AdEle uses an offline multi-objective simulated-annealing-based algorithm (AMOSA) to find an optimal subset of elevators for each router, and an online elevator-selection algorithm to improve elevator selection in the presence of runtime traffic. The following discusses the novel contributions of AdEle. 
\begin{figure}[t]
\centering
\subfigure[]{
\includegraphics[scale=0.43]{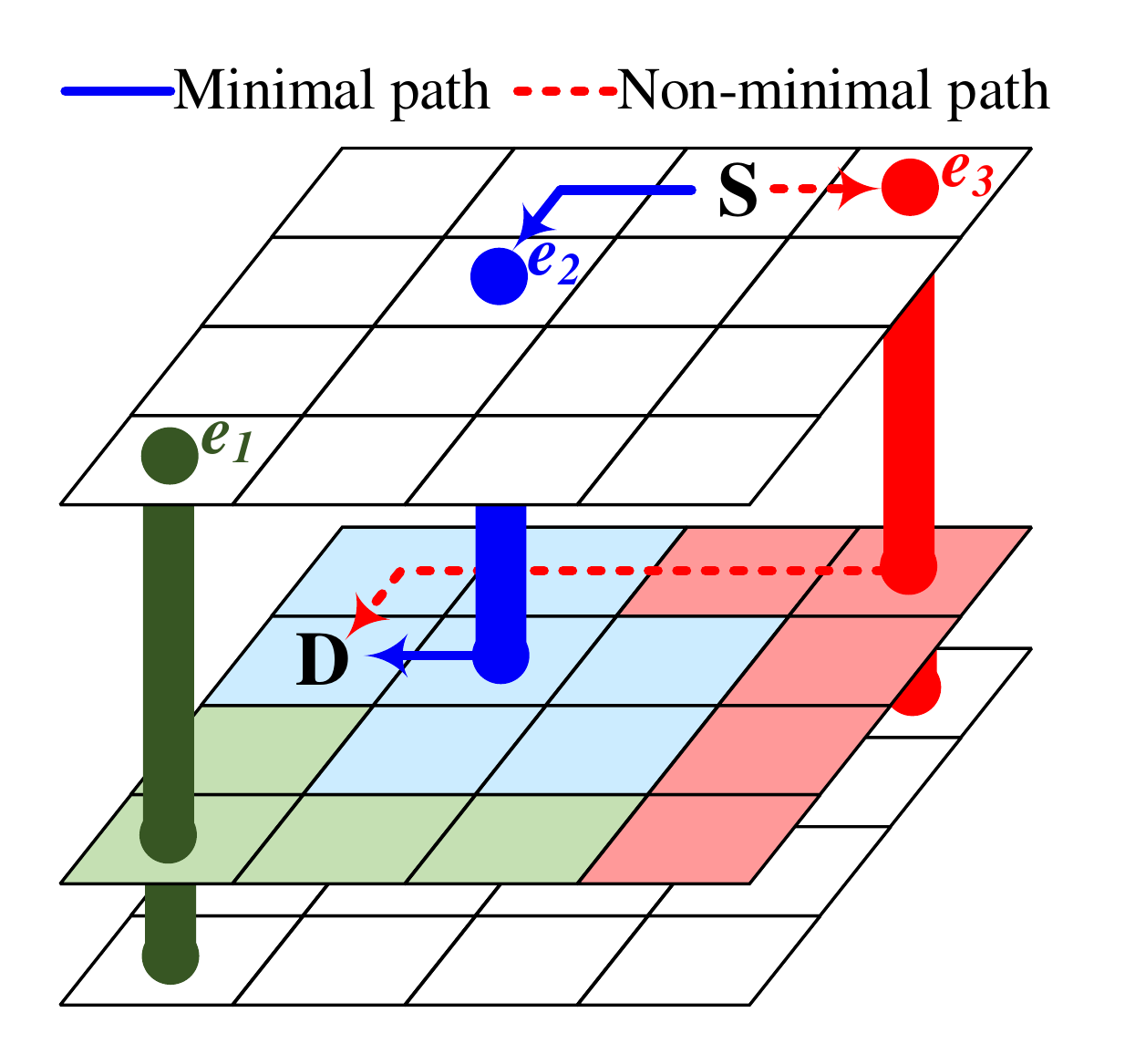}}\label{fig1a}%
\subfigure[]{
\includegraphics[scale=0.53]{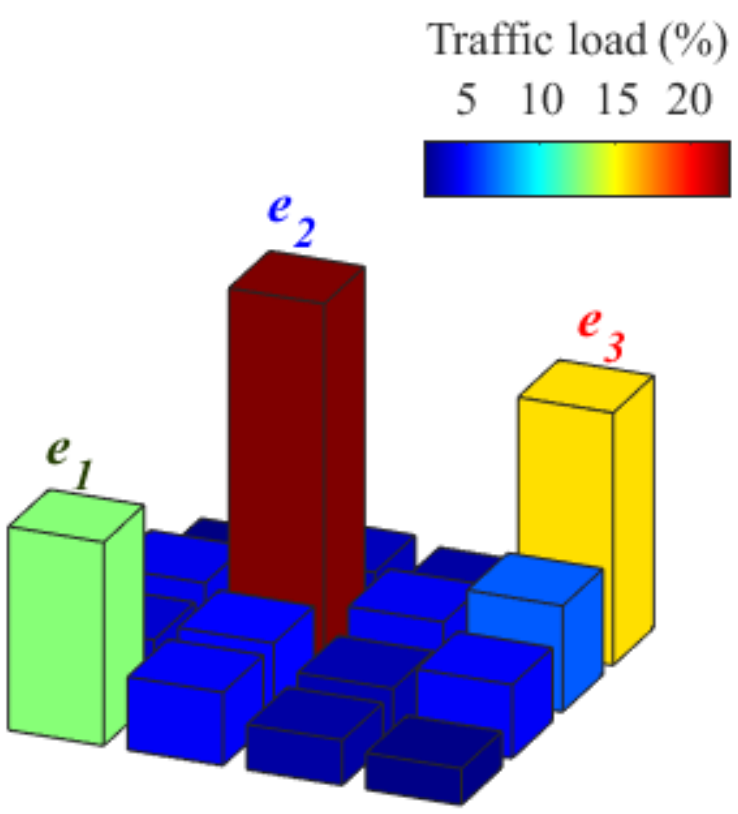}}\label{fig1b}%
\vspace{-0.05in}
\caption{(a) An example PC\=/3DNoC with three elevators ($e_1$, $e_2$, and $e_3$). The routing path from S to D based on Elevator\=/First algorithm~\cite{dubois2011elevator} (dotted-red line) and the minimal path (blue-solid line) are shown. The middle-layer routers are colored based on their Elevator\=/First selected elevator. (b) Traffic load on each router in the middle layer: the $e_2$ elevator is highly congested because of the inefficient elevator selection in Elevator-First algorithm.}
\label{fig1}
\vspace{-0.25in}
\end{figure}

\subsection{Motivation: Routing in PC\=/3DNoCs}
\vspace{-0.1in}
In PC\=/3DNoCs, because of the irregular topology, the routing process requires three main steps: 1) selecting an elevator for each packet in the source router and then routing the packet to that elevator; 2) vertically routing the packet to the destination layer; and 3) routing the packet from the elevator to the destination node. In this routing process, the elevator selection (the first step) is critical as the number of vertical paths (elevators) is much smaller than the number of horizontal paths, putting significantly more traffic pressure on the elevators. %As we will show, AdEle optimizes the elevator selection in the first step to balance the traffic over the elevators, improving network traffic hot-spots.

%The selection is based on closest elevator to source (first elevator)
%To further discuss AdEle's motivation, 
Fig.~\ref{fig1}(a) shows an example of a PC\=/3DNoC with three elevators ($e_1$--$e_3$) using Elevator\=/First-based elevator selection~\cite{dubois2011elevator,taheri2020addressing, coelho2019fl} (i.e., the closest elevator to the source router is selected). Routers are colored with the elevator's color they would use under the Elevator\=/First policy: i.e., four routers will use the green ($e_1$) elevator, seven will use the blue ($e_2$) elevator, and five will use the red ($e_3$) elevator. Unfortunately, such an uneven elevator utilization can put severe traffic pressure on certain elevators ($e_2$ in this example). Ideally, some of the load on the $e_2$ elevator could be assigned to the $e_1$ or $e_3$ elevators, making the $e_2$ elevator less congested. Fig.~\ref{fig1}(b) demonstrates the utilization of the middle-layer routers with Elevator\=/First selection policy under uniform traffic. As can be seen, $e_2$ is highly congested due to the uneven elevator selection. In terms of energy efficiency, the best elevator selection is on the minimal path between the source and destination. However, as can be seen in Fig.~\ref{fig1}(a) for the path between S and D, policies like Elevator\=/First (red-dotted line) do not necessarily choose the minimal path (blue-solid line). 

AdEle will consider both traffic distribution and energy efficiency to select optimal elevators and evenly distribute traffic loads among the elevators. To the best of our knowledge, AdEle is the first congestion- and energy-aware elevator-selection scheme in PC\=/3DNoCs that includes elevator redundancy and online policies to accommodate dynamic traffic behavior while relying only on local router information.  

%\vspace{-0.03in}
\subsection{Optimal Elevator-Subset for Each Router}
%\vspace{-0.05in}
To find the optimal subset of elevators for each router, AdEle performs an offline optimization to distribute the expected traffic load across all elevators and minimize the average inter-node (source to destination) distance. To do this, we first define two optimization objectives: 1) elevator-utilization variance to improve the traffic load distribution, and 2) average inter-node distance to minimize the energy consumption. Leveraging these objective functions, we will use a multi-objective simulated-annealing-based algorithm (AMOSA~\cite{bandyopadhyay2008simulated}) to find the optimal elevator subsets.

\subsubsection{Objective 1 - Elevator Utilization}
To balance the traffic on the elevators, AdEle attempts to minimize the elevator-utilization variance. As discussed above, it is important to evenly distribute the traffic over elevators to avoid highly congested elevators. To calculate the utilization variance, let us consider an $N$-node/router network with a set of elevators $\mathcal{E}=\{e_1,e_2,…,e_E\}$, where $E$ is the total number of elevators. Moreover, assume that during runtime, each router $i$ can select its elevator from a subset $A_i \subseteq \mathcal{E}$. For simplicity, for now we assume that each router selects each elevator from its elevator subset ($A_i$) uniformly (e.g., using a round-robin policy). Therefore, the utilization of elevator $e$ ($U_e$) is:\vspace{-0.05in}
\begin{equation}
U_{e} = \sum_{i=1}^{N}\frac{1}{|A_i|}\sum_{j=1}^{N}f_{ij} \cdot P_{ije},
        \label{Ut_e}\vspace{-0.05in}
\end{equation}
where $f_{ij}$ is the frequency of traffic between routers $i$ and $j$, and $P_{ije}$ denotes whether the routing between routers $i$ and $j$ uses the elevator $e$ ($P=1$) or not ($P=0$). Leveraging \eqref{Ut_e}, the average traffic over all the elevators ($\mu$) is:\vspace{-0.05in} 
\begin{equation}
\mu=\frac{1}{E}\sum_{i=1}^{E}U_{i}.\label{avg_traffic}\vspace{-0.05in}
\end{equation}
Using \eqref{Ut_e} and \eqref{avg_traffic}, elevator-utilization variance is:\vspace{-0.05in}
\begin{equation}
\sigma^2 = \frac{1}{E}\sum_{i=1}^{E}(U_{i}-\mu)^2.
\label{equ_elevator_ut_var}\vspace{-0.05in}
\end{equation}
Minimizing the elevator-utilization variance will result in a better distribution of traffic load on the elevators and lower network latency.

\subsubsection{Objective 2 - Average Distance}
To improve network energy efficiency, AdEle attempts to minimize the average distance. As elevator selection is under consideration here, we only consider inter-layer traffic here. Therefore, the distance between inter-layer nodes $i$ and $j$ over an elevator $e$ can be defined as:
\begin{equation}
D_{ij}^e=\left\{
   \begin{array}{ll}
     0, & \mbox{$i$ and $j$ are on the same layer} \\
     d_{se}+d_e+d_{ed}, & \mbox{otherwise}
   \end{array},
\right.\label{eq:dis}%\vspace{-0.05in}
\end{equation}
where $d_{se}$, $d_{e}$, and $d_{ed}$ are the Manhattan distances between the source and elevator, on the elevator (inter-die), and from the elevator to the destination, respectively. Based on \eqref{eq:dis}, the average inter-layer-node distance in an $L$-layer network is:\vspace{-0.05in}
\begin{equation}
AD = \frac{1}{N \times (\frac{L-1}{L}\times N)}\sum_{i=1}^{N}\frac{1}{|A_i|}\sum_{e=1}^{|A_i|}\sum_{j=1}^{N}D_{ij}^e.
\label{equ_effective_AD}\vspace{-0.05in}
\end{equation}

\subsubsection{Multi-Objective Optimization}
We use a multi-objective simulated annealing-based optimization algorithm (AMOSA~\cite{bandyopadhyay2008simulated}) to find a set of optimal elevator subsets for all the routers in the network ($\mathcal{A}=\{A_1,\ldots,A_N\}$) while minimizing the objective functions in \eqref{equ_elevator_ut_var} and \eqref{equ_effective_AD}. As AMOSA is a multi-objective optimization search, it offers a set of solutions that lie on the Pareto front of the optimization objectives (see \cite{bandyopadhyay2008simulated} for more details). AMOSA-based optimization in AdEle provides different optimal solutions in terms of latency and energy efficiency. From these solutions, a designer can make trade-offs when choosing between more latency-aware or energy-aware solutions (see Fig.~\ref{amosa}). Selection of solutions are discussed in detail in Section~IV.

%\vspace{-0.05in}
\subsection{Adaptive Elevator Selection}
%\vspace{-0.05in}
Here, we discuss how a router $i$ can efficiently select an elevator during runtime from its elevator subset ($A_i$) identified in the previous subsection. As we are interested in an even distribution of traffic load over all the elevators to improve traffic congestion during runtime, we apply an enhanced round-robin (RR) algorithm to select an elevator. In a conventional RR approach, elevators would be selected in a sequential order without considering the runtime traffic. To account for runtime traffic, we include the probability of skipping ($P_{Sik}$) a congested elevator ($k$) for router $i$ in the RR approach. $P_{Sik}$ is adjusted based on the average latency imposed by the elevator $k$, i.e., higher latencies seen using elevator $k$ increases the probability of skipping it in the future. Accordingly, AdEle can adaptively manage dynamic traffic loads and congestion. 

To find $P_{Sik}$, let us first define a cost function associated with making a selection from an elevator subset. After selecting an elevator, AdEle estimates the cost of this selection by considering the time between when the first flit (the header flit) and when the last flit (the tail flit) leave the source router. The latency ($T_{e_k}$) imposed by selecting an elevator $e_k$ from a subset $A_i$ is:\vspace{-0.05in}
\begin{equation}
T_{e_k}= \frac{t_{tail} - t_{head} - l_p}{l_p},
\label{equ_selection_delay}\vspace{-0.05in}
\end{equation}
where $t_{tail}$ and $t_{head}$ denote the time when the tail flit and the header flit leave the source router, respectively. Also, $l_p$ is the length of the packet. The elevator-selection cost ($C_k$) can be updated using the latency of the last selection defined in \eqref{equ_selection_delay} and based on:\vspace{-0.05in}
\begin{equation}
\begin{array}{ll}
C_{k} \leftarrow (a \times T_{e_k})+((1-a)\times C_{k}), &   0 \le a \le 1
\end{array}
\label{equ_Eselection_cost}\vspace{-0.05in}
\end{equation}
where $a$ is a coefficient to increase or decrease the impact of the new cost versus the old cost. We have experimentally found that $a=$~0.2 produces good results in AdEle.

Leveraging \eqref{equ_Eselection_cost}, AdEle can estimate the latency cost at the source router with \textit{only local information}. With wormhole switching, any blocking in an elevator can be propagated along the path from the elevator to the source router. Therefore, blocking at a source router can be interpreted as blocking in the elevator. Note that incorporating global-network information into AdEle would improve the selection policy but be less practical as it will impose high hardware area, energy consumption, and latency costs. 

\begin{table}[t]
\caption{Simulation Setup}\vspace{-0.05in}
\centering
\small
\begin{tabular}{r||l}
\hline
%Simulator & Access Noxim \cite{jheng2010traffic}\\
%\hline
Network size & 4$\times$4$\times$4 and 8$\times$8$\times$4\\
\hline
%Simulation time & 200 K cycles\\
%\hline
%Warm-up time & 10 K cycles\\
%\hline
Routing and VC selection & Elevator\=/First \cite{dubois2011elevator}\\ (w/o elevator selection) & (used to avoid deadlock)\\
\hline
%Switching & Wormhole switching\\
%\hline
Buffer depth & 4 flits \\
\hline
Packet size & 10--30 flits (random) \\
\hline
Traffic pattern & Uniform, Shuffle, and Real\\
\hline
\multicolumn{2}{c}{Elevator-placement patterns}\\
\multicolumn{2}{c}{{ \includegraphics[width=0.24\textwidth, height=0.11\textwidth]{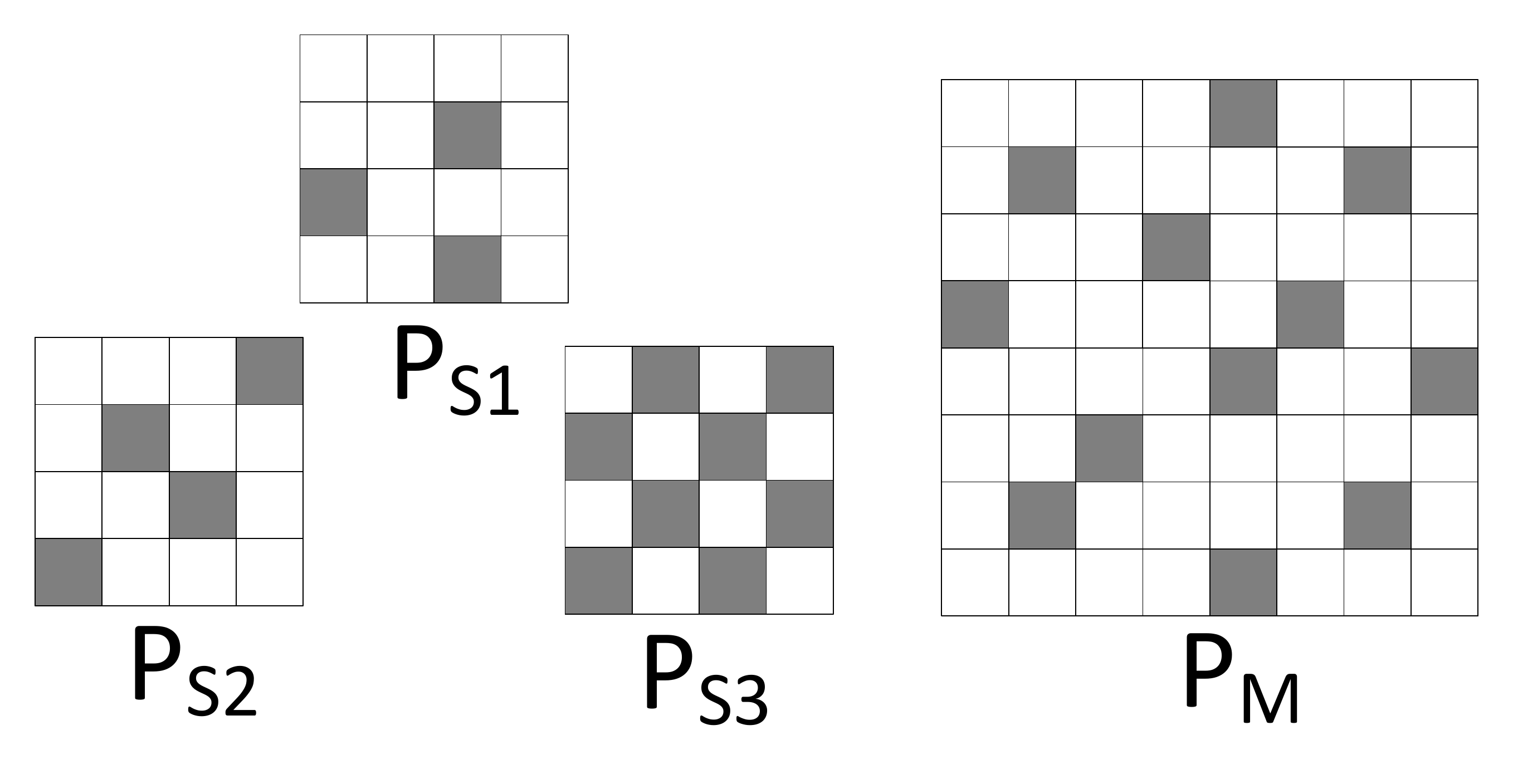}}}\\
\hline
\hline
\end{tabular}
\label{setup}
\vspace{-0.2in}
\end{table}
Considering \eqref{equ_Eselection_cost}, we can define router $i$'s relative cost of selecting elevator $k$ from $A_i$ versus other possible elevators:\vspace{-0.05in}
\begin{equation}
C^{rel}_{ik}= \frac{C_k}{\sum_{p=0}^{|A_i|}C_p}. 
\label{equ_relative_cost}\vspace{-0.05in}
\end{equation}
Based on the relative cost of a particular elevator selection, the possibility of skipping that elevator in the RR approach is:\vspace{-0.05in}
\begin{equation}
P_{Sik}=\left\{
\arraycolsep=1.6pt
   \begin{array}{ll}
     1 - \xi,&\hbox{if $C^{rel}_{ik} \ge \frac{2}{|A_i|}$}\\
     |A_i| \cdot (C^{rel}_{ik}-\frac{1}{|A_i|}) \cdot (1- \xi),& \hbox{if $\frac{2}{|A_i|} > C^{rel}_{ik}\ge \frac{1}{|A_i|}$}\\
     0,&\hbox{otherwise}
   \end{array}
\right.
\end{equation}

Here, $\xi$ is considered to allow for exploring new solutions even under high relative costs ($\xi=$~0.05 in our experiments). To clarify the use of $\xi$, suppose that the $P_{Sik}$ of a selection is 1 because of high congestion. In this case, the elevator $k$ will not be selected in the RR sequence at all and have no chance to update its elevator-selection cost ($C_k$). This would keep $P_{Sik}$ high and prevent the elevator from observing any changes in its cost. To address such an update failure, $\xi$ allows every elevator to be selected with a low probability regardless of $P_{Sik}$ so the cost function has a chance for updating. To improve energy efficiency, when $C_{k}$ is below a threshold for all $k$ (low latency applications) and congestion is not a concern, AdEle will instead choose the elevator along the minimal path (discussed in Section III.A). Here, we experimentally find the threshold that minimizes the latency for each traffic and elevator configuration. Our future work will investigate a dynamic threshold management. \vspace{-0.05in}

%Our future work will consider a dynamic policy.

\pdfoutput=1
\section{Simulation and Evaluation Results}
Here, we compare AdEle against two well-known elevator-selection approaches: Elevator\=/First~\cite{dubois2011elevator} and CDA~\cite{fu2019congestion}. The simulation setup is shown in Table~\ref{setup}. In PC\=/3DNoCs, the number and location of elevators is limited by hardware constraints~\cite{eghbal2015analytical}. Therefore, AdEle is evaluated using different elevator-placement patterns to show that its efficacy is independent of any such patterns. Also, because of performance-area trade-off in PC\=/3DNoCs~\cite{arka2020making}, various elevator concentrations might be employed. Therefore, here we simulate different concentration of elevators to show that AdEle performance is not limited by elevator concentration. Three elevator patterns are considered for a 4$\times$4$\times$4 network ($P_{S1}$--$P_{S3}$) with different levels of elevator concentration. $P_{S1}$ and $P_{S3}$ are extracted to have an optimized average distance and $P_{S2}$ is based on~\cite{coelho2019fl}. A large network (8$\times$8$\times$4) is simulated to show the scalability of AdEle. The pattern for this network ($P_{M}$) is also extracted based on the average distance optimization.
\begin{figure}[t]
\centering
\includegraphics[scale=0.38]{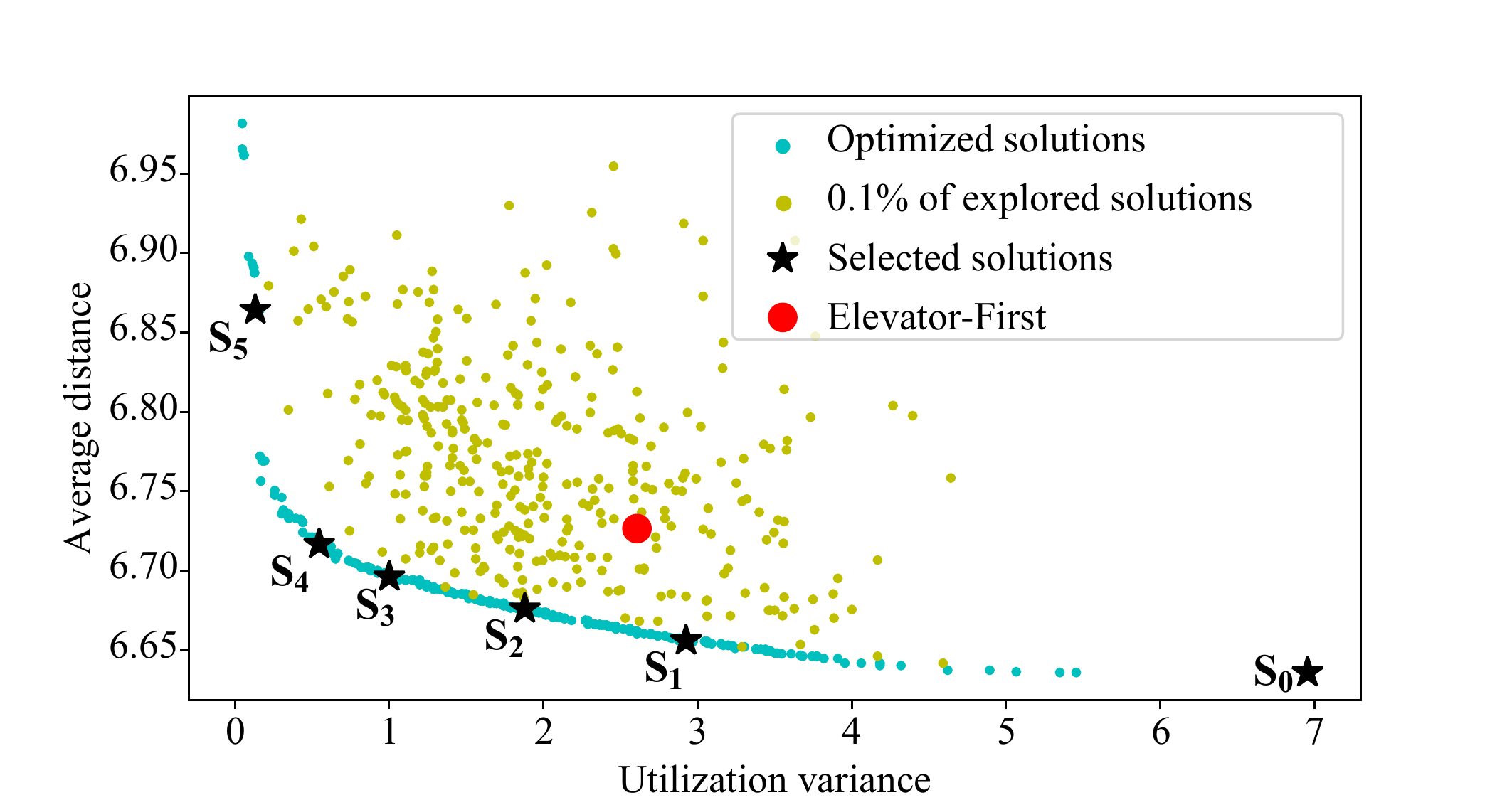}
\vspace{-0.1in}
\caption{Elevator-selection solutions found by AMOSA optimization in AdEle.}
\vspace{-0.15in}
\label{amosa}
\end{figure}
\begin{table}[t]
\caption{Performance of selected solutions from Fig.~\ref{amosa}}
\vspace{-0.05in}
\centering
\small
\begin{tabular}{|c||c|c|c|c|c|c|c|}
\hline
  & Elev.& \multicolumn{6}{c|}{Optimized solutions}\\
  \cline{3-8}
  & First & $S_0$ & $S_1$ & $S_2$ & $S_3$ & $S_4$& $S_5$\checkmark\\
\hline
\hline
%Latency$^*$ & 1 &      3.51&      1.21&  1.07&  0.69 &  0.67 &  0.59\\%Norm Old
Latency$^*$ & 161.4& 396 & 209 & 156.6 & 76.9 &  67.4 & 56.6\\%new
%Latency$^*$ &  1&       2.45 &     1.29 & 0.97 & 0.47 &  0.41 &  0.35\\%Norm new
\hline
%Energy $^\#$ & 1 & 0.97 & 0.99 & 0.99 & 0.99 & 1.01 & 1.03\\%Norm Old
Energy $^\#$ & 94.4 & 93.1 & 94.2 & 94.6 & 94.4 & 94.8 & 98.3\\%new
%Energy$^\#$ &  1 &    0.98 & 0.998 &     1.002  &  1 &    1.004 & 1.04\\%Norm new
\hline
\multicolumn{3}{c}{$^*$Average Latency (cycles)} &  \multicolumn{3}{c}{$^\#$ Energy/flit ($nj$)} & \multicolumn{2}{c}{\checkmark Selected}
\end{tabular}
\vspace{-0.25in}
\label{Amosa_table}
\end{table}

AdEle's offline optimization (see Section~III.B) is implemented in Python to extract the elevator subsets for routers. These subsets are added to the AdEle router implemented in Access Noxim simulator~\cite{jheng2010traffic}. We considered uniform traffic for the offline optimization, the most pessimistic assumption (i.e., traffic is not known \textit{a priori}), while the network simulations are done using different synthetic and real-application traffics. Our analysis will demonstrate that AdEle does not require runtime traffic in its offline optimization as its online selection policy will adjust to runtime traffic. However, AdEle can use the runtime traffic during elevator-subset selection to offer further latency and energy improvement.\vspace{-0.05in} 

\subsection{AMOSA Elevator-Subset Exploration}\vspace{-0.05in} 
As discussed in Section~III, AMOSA finds various solutions with different latency and energy-efficiency. To show the solution selection process, the optimization for $P_{M}$ is detailed here. A small sample of AMOSA's explored solutions is shown in Fig.~\ref{amosa}. As AMOSA explores the solution space, it makes its way towards the Pareto front (blue curve) to find the optimal trade-offs between utilization variance and average distance. Given the set of solutions, depending on the importance of energy efficiency (average distance) and latency (utilization variance), the final solution can be selected. For brevity, several of these points spread along the Pareto front are selected for network simulation ($S_0$ to $S_5$) where the results are summarized in Table~\ref{Amosa_table}. Considering Table~\ref{Amosa_table} and Fig.~\ref{amosa}, lower utilization variance and lower average distance improves the latency and energy consumption, respectively. As we are able to significantly reduce the latency with fairly minimal increases in energy, we select $S_5$ for further analysis. Similarly, we  select the solution for $P_{S1}$--$P_{S3}$.
\begin{figure*}[t]
\centering
\subfigure[$P_{S1}$-Uniform]{\includegraphics[width=.225\textwidth]{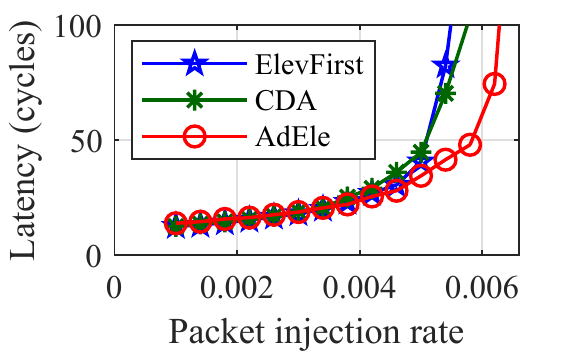}}\hfill\vspace{-0.1in}
\subfigure[$P_{S2}$-Uniform]{\includegraphics[width=.225\textwidth]{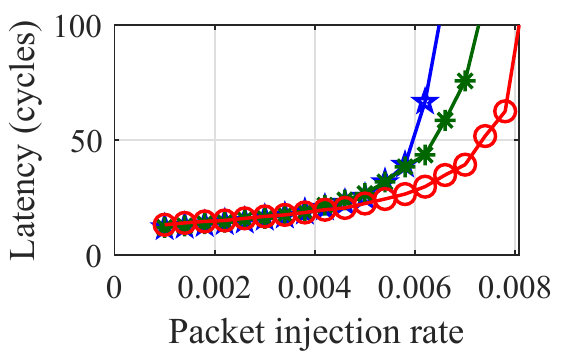}}\hfill
\subfigure[$P_{S3}$-Uniform]{\includegraphics[width=.225\textwidth]{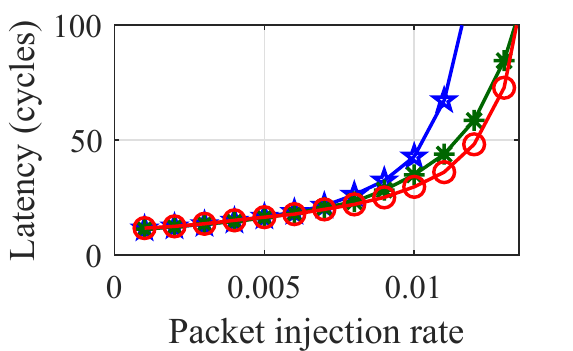}}\hfill
\subfigure[$P_{M}$-Uniform]{\includegraphics[width=.225\textwidth]{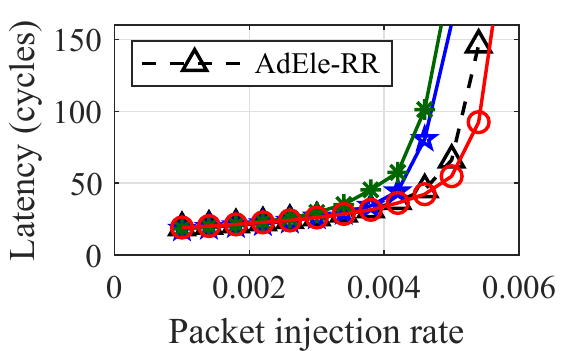}}\hfill
\subfigure[$P_{S1}$-Shuffle]{\includegraphics[width=.225\textwidth]{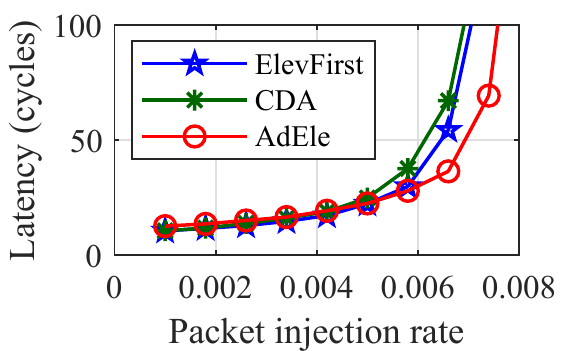}}\hfill
\subfigure[$P_{S2}$-Shuffle]{\includegraphics[width=.225\textwidth]{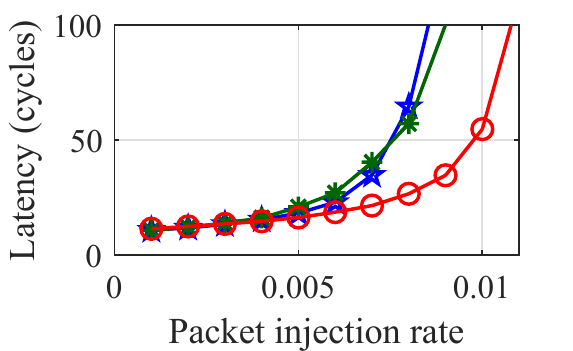}}\hfill
\subfigure[$P_{S3}$-Shuffle]{\includegraphics[width=.225\textwidth]{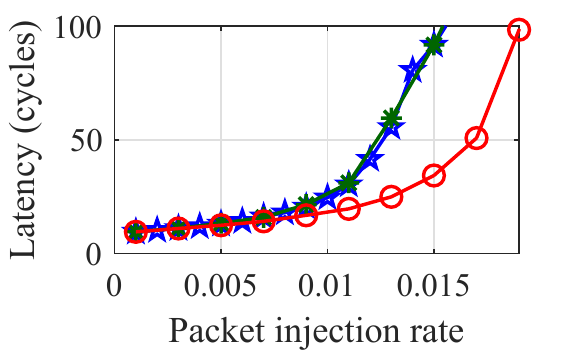}}\hfill
\subfigure[$P_{M}$-Shuffle]{\includegraphics[width=.225\textwidth]{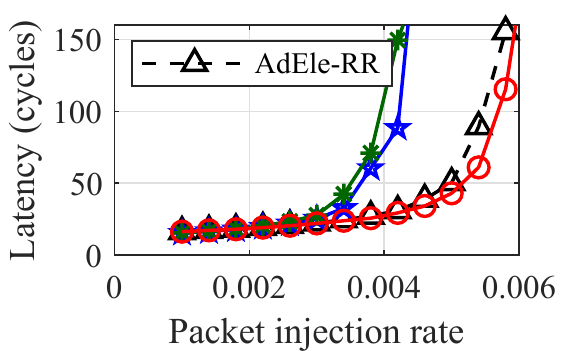}}
\vspace{-0.1in}
\caption{Average latency for Elevator\=/First, CDA, and AdEle under uniform (a--d) and shuffle (e--h) traffic and with different elevator-placement patterns.}
\label{fig_synthetic}
\vspace{-0.2in}
\end{figure*}
\begin{figure}[t]
\centering
\includegraphics[scale=0.6]{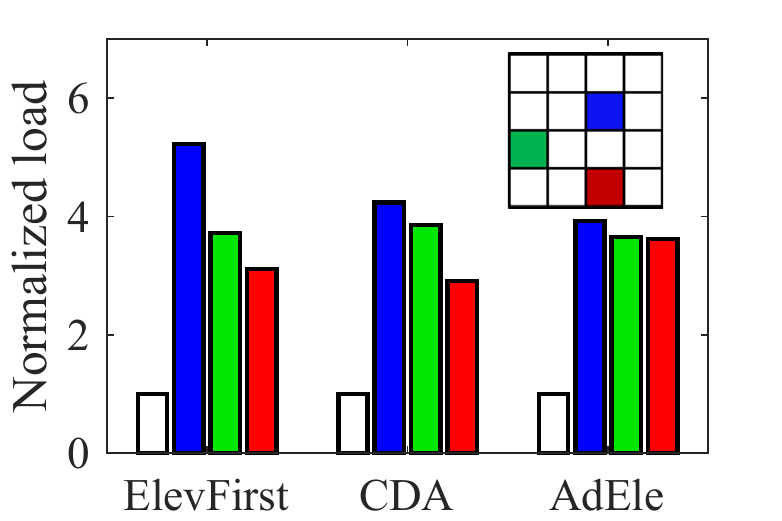}
\vspace{-0.1in}
\caption{Traffic load over routers with elevators (blue, green, and red) normalized to the average load over routers without an elevator (white bar).}
\vspace{-0.25in}
\label{TrafficDistribute}
\end{figure}

\vspace{-0.05in}
\subsection{AdEle Performance Under Synthetic Traffic}\vspace{-0.05in}
To compare AdEle with Elevator\=/First and CDA, we first evaluate the average latency under uniform and shuffle traffic patterns and with different elevator-placement patterns in Fig.~\ref{fig_synthetic}. Across all the traffic and elevator-placement patterns, AdEle achieves the lowest latency and highest saturation threshold. Note that CDA is able to approach AdEle's performance because it considers global intra-layer traffic. In this work, we do not consider the high cost of CDA's global information sharing and optimistically assume that the information is instantaneously received at every router. In reality, CDA will likely perform much worse with stale information or include significant implementation overhead. With a higher elevator density (e.g., $P_{S3}$), the elevator congestion issue is less critical and intra-layer traffic will be more critical. Similarly, in a network with larger horizontal dimensions like $P_{M}$, intra-layer traffic is more important. Yet, AdEle shows better performance even with a high density of elevators and for $P_M$. Recall that AdEle's offline optimization step used uniform traffic. Yet, as Figs.~\ref{fig_synthetic}(e)--(h) show, while the traffic is new for AdEle, it still achieves the lowest latency because its online selection policy can monitor runtime congestion and select better elevators. If the traffic is known \textit{a priori}, AdEle can use this traffic information during offline optimization to improve elevator-subset selection even further. For $P_{M}$ in Figs.~\ref{fig_synthetic}(d) and~\ref{fig_synthetic}(h), we also include the average latency of AdEle with standard RR selection. This demonstrates that AdEle's proposed online skipping policy achieves higher improvements in latency compared to RR in both uniform and shuffle traffic patterns.

\begin{figure}[t]
\centering
\subfigure[Low injection rate]{\includegraphics[scale=0.62]{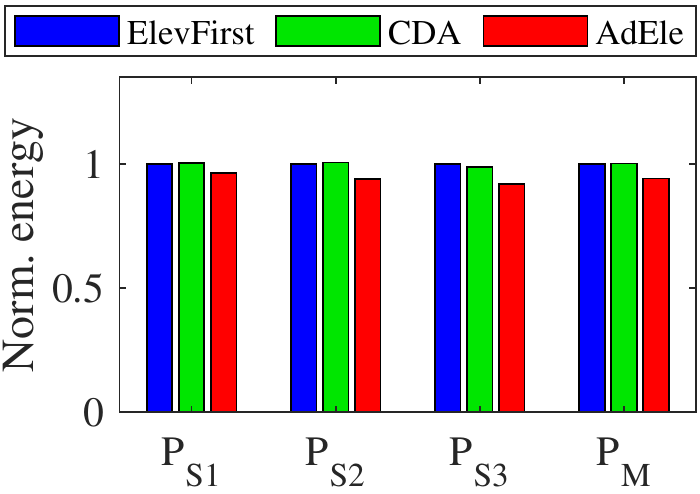}}\hspace{-0.0em}%
\subfigure[High injection rate]{\includegraphics[scale=0.605]{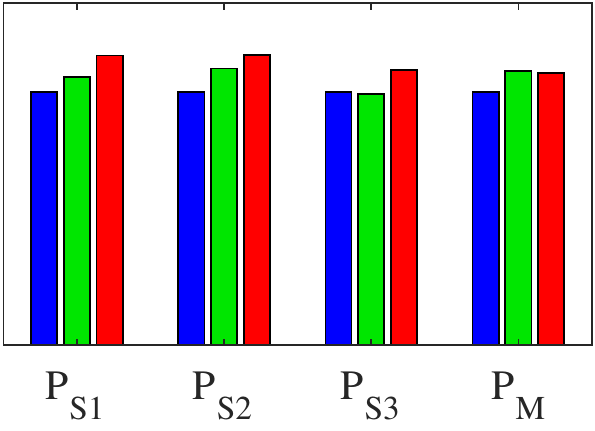}}\hspace{-0.2em}%
\vspace{-0.07in}
\caption{Energy per flit for Elevator\=/First (ElevFirst), CDA, and AdEle normalized to ElevFirst, and under different injection rates.}
\label{energy}
\vspace{-0.25in}
\end{figure}

To show the main reason for latency improvement when using AdEle, the load distribution over routers with elevators for $P_{S1}$ is shown in~Fig.~\ref{TrafficDistribute}. The white bar shows the average load over elevator-less routers. The other colored bars show the load over different elevators. As can be seen, AdEle reduces the load on the highest utilized elevator (blue elevator). The energy consumption for each approach and elevator placement is shown in Fig.~\ref{energy} for low (1E$-$3) and high (5E$-$3 to 1.2E$-$2) injection rates based on the saturation point (injection rate at which latency is 10$\times$ zero-load latency) for each configuration. For low injection rates, AdEle has the lowest energy consumption because it switches to minimal routing and uses the minimal paths. On the other hand, AdEle incurs a small energy overhead (less than 9.7\% compared to CDA) under high injection rates to take non-minimal paths and improve traffic congestion. If less energy overhead is desired, AdEle can use configurations with lower energy (see Table~\ref{Amosa_table}). 

\begin{figure*}[t]
\centering
\subfigure[$P_{S1}$]{\includegraphics[scale=0.57]{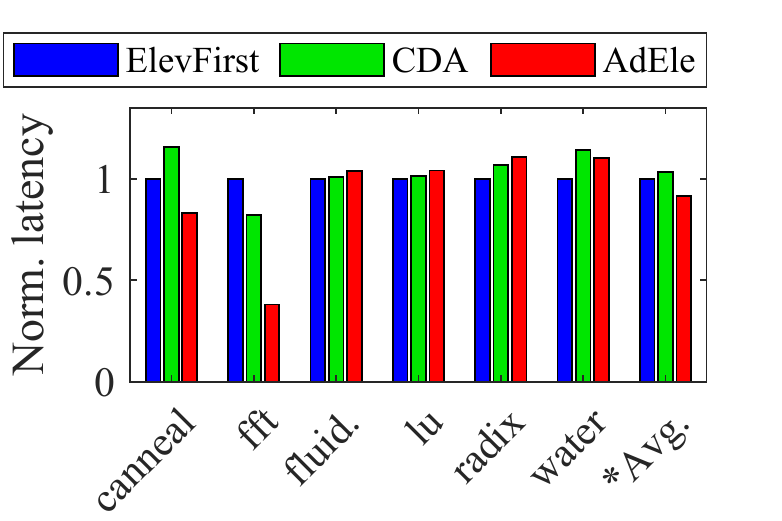}}\hspace{-0.2em}%
\subfigure[$P_{S2}$]{\includegraphics[scale=0.57]{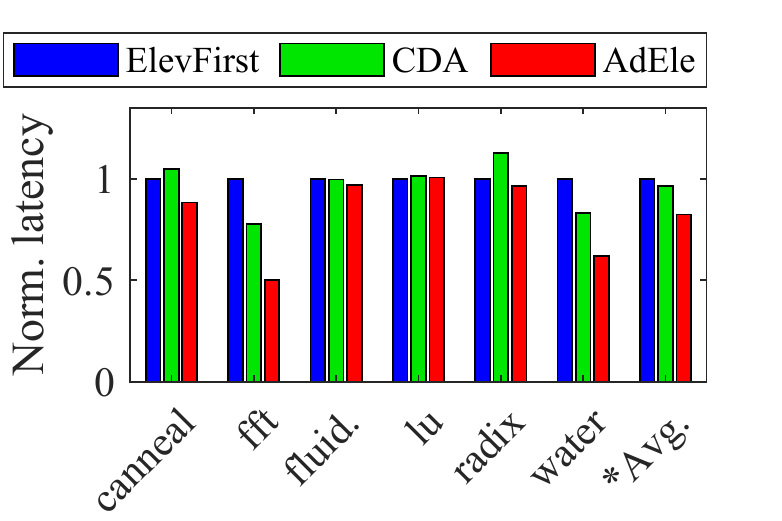}}\hspace{-0.2em}%
\subfigure[$P_{S3}$]{\includegraphics[scale=0.57]{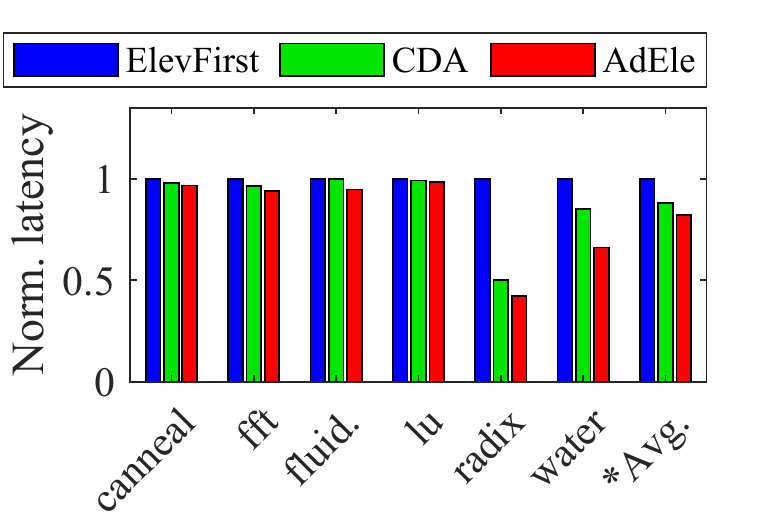}}\hspace{-0.2em}%
\subfigure[Energy: $P_{S1}$--$P_{S3}$]{\includegraphics[scale=0.57]{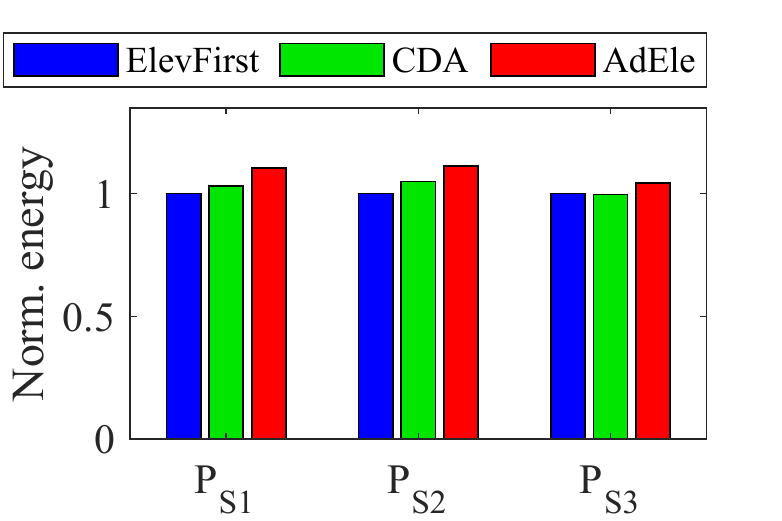}}\hspace{-0.2em}%
\vspace{-0.1in}
\caption{Latency ((a)--(c), per application) and energy ((d), averaged across all applications) for Elevator\=/First (ElevFirst), CDA, and AdEle normalized to ElevFirst under real-application traffic with different elevator-placement patterns. $^*$Avg. in (a)--(c) is the average of all six applications.}
\label{fig_real}
\vspace{-0.24in}
\end{figure*}

%\vspace{-0.05in}
\subsection{AdEle Performance under Real-Application Traffic}
We extracted the traffic of several SPLASH-2~\cite{woo1995splash} and PARSEC~\cite{bienia2011benchmarking} benchmarks using Gem5~\cite{binkert2011gem5} for real-application simulations. Because Gem5 is limited to 64 cores, we demonstrate our results for $P_{S1}$--$P_{S3}$. As shown in Figs.~\ref{fig_real}(a)--(c), AdEle improves the network latency in nearly all cases. In particular, AdEle has more improvements in applications with higher traffic loads (canneal, fft, radix, and water) as there is more opportunity to reduce the resulting elevator congestion. In applications with lower traffic loads (fluidanimate and lu), AdEle maintains similar performance to the other approaches as there is little contention on the elevators and the latency is close to zero-load latency. Although $P_{S1}$ still shows some improvements for AdEle, the lower number of elevators (three) results in minimal opportunity for AdEle to redirect traffic and improve latency. On average, AdEle improves the network latency by 10.9\% (up to 14.6\%) compared to CDA and by 14.6\% (up to 18\%) compared to Elevator\=/First under $P_{S1}$--$P_{S3}$. Fig.~\ref{fig_real}(d) shows, for each elevator-placement pattern ($P_{S1}$--$P_{S3}$), the average energy over all the applications normalized to Elevator\=/First. AdEle imposes a small overhead because it may route packets over non-minimal paths in case of congestion to improve latency. Compared to CDA, AdEle has on average 6.9\%, 6.2\% and 4.8\% energy overhead under $P_{S1}$, $P_{S2}$, and $P_{S3}$, respectively. 

\vspace{-0.05in}
\subsection{Hardware-Area Analysis and Comparison}
\label{hardware_analysis_section}
Routers' hardware of Elevator\=/First, AdEle, and CDA are implemented and analyzed using Cadence Genus in 45~nm technology. Here, we consider a 1~GHz clock. The results are shown in Table~\ref{hardware}. Compared to CDA$^*$, AdEle has a smaller area overhead. This is because AdEle only requires local traffic information while CDA requires a table to save global traffic information and find the best path in each router. However, CDA$^*$'s area overhead is an optimistic assumption here as it does not include any overhead related to the actual sharing of information. Therefore, real CDA will likely impose higher area and latency overheads. Also, AdEle does not affect the router stages and will scale well with the network size, while CDA requires an additional cycle (or more for larger networks) to update its tables.\vspace{-0.05in}
\begin{table}[t!]
\caption{Area analysis}
\vspace{-0.05in}
\centering
\small
\begin{tabular}{|c||c|c|c}
\cline{1-3}
   & Cycles & Router area $(\mu m^2)$ & \\
\cline{1-4}
\cline{1-4}
Base (ElevFirst) & 1 & 35550 & \multicolumn{1}{c|}{Overhead}\\
\hline
%CDA$^*$  & $37619$ & \multicolumn{1}{c|}{$5.8\%$} \\
CDA$^*$  & 2 & 41088 & \multicolumn{1}{c|}{14.4\%} \\
\hline
AdEle & 1 & 36640 & \multicolumn{1}{c|}{3.1\%} \\
\hline
\multicolumn{4}{c}{$^*$global information sharing is not included.} 
\end{tabular}
\label{hardware}
\vspace{-0.26in}
\end{table}
\pdfoutput=1
\section{Conclusion}
This paper proposes AdEle, as adaptive congestion- and energy-aware elevator-selection scheme to address elevator overutilization in partially connected 3D NoCs. Employing a set of elevators instead of one elevator for each source router, AdEle monitors the network traffic and provides an online policy to select the proper elevator while considering runtime traffic loads. AdEle only requires local router information and is able to improve average latency in various scenarios under both synthetic and real traffic at the cost of less than 6.9\% in energy consumption. Moreover, AdEle can be easily adjusted to consider faults, which is of great interest in PC\=/3DNoCs, while considering elevator congestion.

%This is just for information (please don't remove)--energy overhead in comparison with CDA in high injection rate under synthetic traffic: PS1: 8.1%  PS2: 4.2%   PS3: 8.2%   PM:1%
%Energy overhead in comparison with CDA in real application average  10.5%, 10.3% and 10% overheads for PS1, PS2 and PS3

\section*{Acknowledgment}
This work was supported by the National Science Foundation (NSF) under grant number CNS-2046226.

\bibliographystyle{IEEEtran}
\bibliography{ref}

% \begin{thebibliography}{00}
% \bibitem{b1} G. Eason, B. Noble, and I. N. Sneddon, ``On certain integrals of Lipschitz-Hankel type involving products of Bessel functions,'' Phil. Trans. Roy. Soc. London, vol. A247, pp. 529--551, April 1955.
% \bibitem{b2} J. Clerk Maxwell, A Treatise on Electricity and Magnetism, 3rd ed., vol. 2. Oxford: Clarendon, 1892, pp.68--73.
% \bibitem{b3} I. S. Jacobs and C. P. Bean, ``Fine particles, thin films and exchange anisotropy,'' in Magnetism, vol. III, G. T. Rado and H. Suhl, Eds. New York: Academic, 1963, pp. 271--350.
% \bibitem{b4} K. Elissa, ``Title of paper if known,'' unpublished.
% \bibitem{b5} R. Nicole, ``Title of paper with only first word capitalized,'' J. Name Stand. Abbrev., in press.
% \bibitem{b6} Y. Yorozu, M. Hirano, K. Oka, and Y. Tagawa, ``Electron spectroscopy studies on magneto-optical media and plastic substrate interface,'' IEEE Transl. J. Magn. Japan, vol. 2, pp. 740--741, August 1987 [Digests 9th Annual Conf. Magnetics Japan, p. 301, 1982].
% \bibitem{b7} M. Young, The Technical Writer's Handbook. Mill Valley, CA: University Science, 1989.
% \end{thebibliography}
% \vspace{12pt}
\end{document}